\documentclass[12pt,aps,showpacs]{revtex4}
\usepackage{amsmath}
\usepackage{amssymb}
\usepackage{amsfonts}

\input{tcilatex}
\begin{document}

\title{Time Fractional Schr\"{o}dinger Equation; \\
Fox's H-functions and the Effective Potential}
\author{Sel\c{c}uk \c{S}. Bay\i n}
\affiliation{Middle East Technical University, Institute of Applied Mathematics, Ankara,
TURKEY 06800 
\email{bayin@metu.edu.tr}
}
\date{\today }

\begin{abstract}
After introducing the formalism of the general space and time fractional Schr%
\"{o}dinger equation,\ we concentrate on the time fractional\ Schr\"{o}%
dinger equation and present new results via the elegant language of Fox's
H-functions. We show that the general time dependent part of the wave
function for the separable solutions of the time fractional Schr\"{o}dinger
equation is the Mittag-Leffler function with an imaginary argument by two
different methods. After separating the Mittag-Leffler function into its
real and imaginary parts, in contrast to existing works, we show that the
total probability is $\leq 1$ and decays with time. Introducing the
effective potential approach, we also write the Mittag-Leffler function with
an imaginary argument as the product of its purely decaying and purely
oscillating parts. In the light of these, we reconsider the simple box
problem.
\end{abstract}

\maketitle

\section{Introduction}

The history of fractional calculus can be traced all the way back to
Leibniz. In a letter to L'Hopital (1695), Leibniz mentions that he has an
expression that looks like a fractional derivative [1, 2]. Later, Euler
notices that using his gamma function fractional integrals may be possible.
However, a systematic development of the subject does not come until
nineteenth century, where Riemann, Liouville, Gr\"{u}nwald and Letnikov play
key roles [1-6]. On the application side, after it became clear that
anomalous diffusion can be investigated in terms of the fractional diffusion
equation, the situation has been changing rapidly [1-6]. The first
applications to quantum mechanics started with Laskin in 2000, where he
considered the path integral formulation of quantum mechanics over L\'{e}vy
paths and showed that the corresponding equation of motion is the space
fractional Schr\"{o}dinger equation [7-17]. In 2004, Naber [18] discussed
the time fractional Schr\"{o}dinger equation [18-24] and obtained the time
dependent part of the wave function for the separable solutions as the
Mittag-Leffler function with an imaginary argument, which he wrote as a
purely oscillatory term plus an integral that can not be evaluated
analytically by standard techniques [18]. The main arguments and the
conclusions of the paper were based on the \textit{critical} assumption that
this integral is uniformly decaying, hence could be neglected for
sufficiently large times. This affected the main conclusions of the paper by
giving probabilities greater than $1.$ Later, others have also followed this 
\textit{critical} assumption [22-24].

We evaluate this integral exactly by using the elegant language of Fox's $H-$%
functions [25-27] and show that the decomposition of the Mittag-Leffler
function as the sum of its purely oscillating and purely decaying parts is
not possible. Since this point implies dramatic changes in the conclusions
of the previous works [18, 22-24], we reconsider the problem of the time
fractional Schr\"{o}dinger equation.

After introducing the time and space fractional Schr\"{o}dinger equation in
Section II, in Section III we concentrate on the time fractional Schr\"{o}%
dinger equation and discuss its separable solutions. We show that the time
dependence is given as the Mittag-Leffler function with an imaginary
argument by two different methods. After separating the Mittag-Leffler
function into its real and imaginary parts, we show that the total
probability is always $\leq 1$ and decays with time. We also discuss the
asymptotic forms of the wave function. In Section IV, we discuss the simple
box problem, and in Section V, we introduce the effective potential
approach. Using the effective potential, in Section VI we show that the
Mittag-Leffler function with an imaginary argument can be written, not as
the sum, but as the product of its purely decaying and purely oscillating
parts. In Section VII, we discuss our results. \ \ \ \ 

\section{Fractional Schr\"{o}dinger Equation}

To write the fractional Schr\"{o}dinger equation\textbf{,} we use the fact
that the Schr\"{o}dinger equation is analytic in the lower half complex $t-$%
plane, and perform a Wick rotation, $t\rightarrow -it,$ on the one
dimensional Schr\"{o}dinger equation, 
\begin{equation}
i\hbar \frac{\partial \Psi }{\partial t}=-\frac{\hbar ^{2}}{2m}\frac{%
\partial ^{2}\Psi }{\partial x^{2}}+V(x)\Psi (x,t),
\end{equation}%
to write 
\begin{equation}
-\hbar \frac{\partial \Psi }{\partial t}=-\frac{1}{2m}\left( \hbar \frac{%
\partial }{\partial x}\right) ^{2}\Psi +V(x)\Psi (x,t).
\end{equation}%
This is nothing but the Bloch equation [6] 
\begin{equation}
\frac{\partial \Psi }{\partial t}=\text{ }\check{D}\frac{\partial ^{2}\Psi }{%
\partial x^{2}}-\frac{1}{\hbar }V(x)\Psi (x,t),\text{ }\check{D}>0,
\end{equation}%
where $\check{D}=\dfrac{\hbar }{2m}$ is the quantum diffusion constant and $%
V(x)$ is the potential. We can now write the time and space fractional
version of the Bloch equation as 
\begin{equation}
_{0}^{C}D_{t}^{\alpha }\Psi (x,t)=\frac{1}{\hbar }\check{D}_{\alpha ,\beta
}\hbar ^{\beta }R_{x}^{\beta }\Psi (x,t)-\frac{1}{\hbar }V(x)\Psi (x,t),%
\text{ }0<\alpha <1,\text{ }1<\beta <2,
\end{equation}%
where $_{0}^{C}D_{t}^{\alpha }$ is the Caputo derivative and $R_{x}^{\beta }$
is the Riesz derivative (Appendix A). When there is no room for confusion,
we also write $_{0}^{C}D_{t}^{\alpha }$ $\equiv \frac{\partial ^{\alpha }}{%
\partial t^{\alpha }}$ and $R_{x}^{\beta }$ $\equiv \nabla ^{\beta }\equiv 
\frac{\partial ^{\beta }}{\partial x^{\beta }}$. We have also introduced a
new quantum diffusion constant, $\check{D}_{\alpha ,\beta },$ with the
appropriate units. Performing an inverse Wick rotation, $t\rightarrow it,$
we obtain the most general fractional version of the Schr\"{o}dinger
equation as 
\begin{equation}
\frac{\partial ^{\alpha }}{\partial t^{\alpha }}\Psi (x,t)=\frac{i^{\alpha }%
}{\hbar }\check{D}_{\alpha ,\beta }\left( \hbar \frac{\partial }{\partial x}%
\right) ^{\beta }\Psi (x,t)-\frac{i^{\alpha }}{\hbar }V(x)\Psi (x,t).
\end{equation}

\subsection{Space Fractional Schr\"{o}dinger Equation}

When $\alpha=1,$ Equation (5) becomes the space fractional Schr\"{o}dinger
equation investigated by Laskin [7]: 
\begin{equation}
\frac{\partial}{\partial t}\Psi(x,t)=\frac{i}{\hbar}\check{D}%
_{1,\beta}\left( \hbar\frac{\partial}{\partial x}\right) ^{\beta}\Psi(x,t)-%
\frac{i}{\hbar }V(x)\Psi(x,t),
\end{equation}
which he wrote as 
\begin{equation}
i\hbar\frac{\partial}{\partial t}\Psi(x,t)=-D_{\beta}\left[ \hbar \nabla%
\right] ^{\beta}\Psi(x,t)+V(x)\Psi(x,t).
\end{equation}
Laskin called $\left[ \hbar\nabla\right] ^{\beta}$ the quantum Riesz
derivative and $D_{\beta}$ in the above equation is the quantum diffusion
constant for the space fractional Schr\"{o}dinger equation, where as $%
\beta\rightarrow2,$ $D_{\beta}\rightarrow1/2m.$

\subsection{Time Fractional Schr\"{o}dinger Equation}

For $\beta =2,$ we get 
\begin{equation}
\frac{\partial ^{\alpha }}{\partial t^{\alpha }}\Psi (x,t)=\frac{i^{\alpha }%
}{\hbar }\check{D}_{\alpha ,2}\left( \hbar \frac{\partial }{\partial x}%
\right) ^{2}\Psi (x,t)-\frac{i^{\alpha }}{\hbar }V(x)\Psi (x,t),
\end{equation}%
or 
\begin{equation}
\frac{\partial ^{\alpha }}{\partial t^{\alpha }}\Psi (x,t)=i^{\alpha }\check{%
D}_{\alpha ,2}\hbar \frac{\partial ^{2}}{\partial x^{2}}\Psi (x,t)-\frac{%
i^{\alpha }}{\hbar }V(x)\Psi (x,t).
\end{equation}%
By defining a new quantum diffusion constant, $D_{\alpha }=\check{D}_{\alpha
,2}\hbar ,$ we can write the time fractional Schr\"{o}dinger equation as 
\begin{equation}
\frac{\partial ^{\alpha }}{\partial t^{\alpha }}\Psi (x,t)=i^{\alpha
}D_{\alpha }\frac{\partial ^{2}}{\partial x^{2}}\Psi (x,t)-\frac{i^{\alpha }%
}{\hbar }V(x)\Psi (x,t),\text{ }0<\alpha <1,
\end{equation}%
where $D_{\alpha }\rightarrow \hbar /2m$ as $\alpha \rightarrow 1.$ Note
that there is a certain amount of ambiguity in writing the time fractional
version of the Schr\"{o}dinger equation [18]. We will address this point
again in Section VII.

\section{Separable Solutions of the Time Fractional Schr\"{o}dinger Equation}

We now reconsider the time fractional Schr\"{o}dinger equation. Assuming a
separable solution of the form $\Psi (x,t)=X(x)T(t),$ we obtain the
equations to be solved for $X(x)$ and $T(t),$ respectively, as 
\begin{align}
D_{\alpha }\frac{d^{2}X(x)}{dx^{2}}-\frac{V(x)}{\hbar }X(x)& =\lambda
_{n}X(x), \\
\frac{\partial ^{\alpha }T(t)}{\partial t^{\alpha }}& =i^{\alpha }\lambda
_{n}T(t),\text{ }0<\alpha <1,
\end{align}%
where the spatial equation [Eq. (11)] has to be solved with the appropriate
boundary conditions, and $\lambda _{n}$ is the separation constant. At this
point, the index $n$ is redundant but we keep it for the cases where $%
\lambda $ is discrete. Since in the limit as $\alpha \rightarrow 1,$ $%
D_{\alpha }\rightarrow \hbar /2m$ and $\lambda _{n}\rightarrow -E_{n}/\hbar $%
, for physically interesting cases we will take $\lambda _{n}<0.$

\subsection{Time Dependence $-$ Method I}

Using Equation (A7) in Appendix A, we take the Laplace transform of Equation
(12): 
\begin{equation}
\pounds \left\{ \frac{\partial ^{\alpha }T(t)}{\partial t^{\alpha }}\right\}
=i^{\alpha }\lambda _{n}\pounds \left\{ T(t)\right\} ,
\end{equation}%
to write 
\begin{equation}
s^{\alpha }\widetilde{T}(s)-s^{\alpha -1}T(0)=i^{\alpha }\lambda _{n}%
\widetilde{T}(s),
\end{equation}%
where $\widetilde{T}(s)=\pounds \left\{ T(t)\right\} .$ This gives the
Laplace transform of the time dependence of the wave function as 
\begin{align}
\widetilde{T}(s)& =T(0)\frac{s^{\alpha -1}}{s^{\alpha }-i^{\alpha }\lambda
_{n}}\  \\
& =T(0)\frac{s^{-1}}{1-i^{\alpha }\lambda _{n}s^{-\alpha }}.
\end{align}%
Using geometric series, we can also write this as 
\begin{align}
\widetilde{T}(s)& =T(0)\sum_{m=0}^{\infty }\left( i^{\alpha }\lambda
_{n}s^{-\alpha }\right) ^{m}s^{-1} \\
& =T(0)\sum_{m=0}^{\infty }i^{\alpha m}\lambda _{n}^{m}s^{-m\alpha -1},
\end{align}%
which converges for $\left\vert i^{\alpha }\lambda _{n}s^{-\alpha
}\right\vert <1.$ The inverse Laplace transform of this can be found easily
to yield the time dependent part of the wave function as 
\begin{align}
T(t)& =T(0)\sum_{m=0}^{\infty }\frac{i^{\alpha m}\lambda _{n}^{m}t^{\alpha m}%
}{\Gamma (1+\alpha m)} \\
& =T(0)\sum_{m=0}^{\infty }\frac{\left( i^{\alpha }\lambda _{n}t^{\alpha
}\right) ^{m}}{\Gamma (1+\alpha m)} \\
& =T(0)E_{\alpha }(i^{\alpha }\lambda _{n}t^{\alpha }),
\end{align}%
where $E_{\alpha }(i^{\alpha }\lambda _{n}t^{\alpha })$ is the
Mittag-Leffler function with an imaginary argument.

\subsection{Time Dependence $-$ Method II}

To find the time dependence of the wave function, an alternate path can also
be taken by directly evaluating the integral [18] 
\begin{equation}
T(t)=T(0)\left[ \frac{\ 1}{2\pi i}\int_{\gamma-i\infty}^{\gamma+i\infty}%
\frac{e^{st}s^{\alpha-1}ds}{s^{\alpha}-i^{\alpha}\lambda_{n}}\right] .
\end{equation}
The integrand has a pole at $s_{1}=0$ due to the numerator, and a pole at $%
s_{2}=i\lambda_{n}^{1/\alpha}$ due to the denominator, hence can be
rewritten as (Bayin [6] and supplements of Ch. 14, [18]) 
\begin{equation}
T(t)=T(0)\left[ \frac{e^{i\lambda_{n}^{1/\alpha}t}}{\alpha}-\frac{\sigma
\sin\alpha\pi}{\pi}\int_{0}^{\infty}\frac{e^{-xt}x^{\alpha-1}dx}{x^{2\alpha
}-2\sigma\cos\alpha\pi\text{ }x^{\alpha}+\sigma^{2}}\right] \ .
\end{equation}
For $\lambda_{n}>0,$ the above expression is valid for all $0<\alpha<1.$ For 
$\lambda_{n}<0,$ the second pole, $s_{2}=\left\vert \lambda_{n}\right\vert
^{1/\alpha}e^{i(\pi/\alpha+\pi/2)},$ has both real and imaginary parts.
Since the branch cut is located along the negative real axis, we have to
exclude the $\alpha=\frac{2}{5},\frac{2}{9},\frac{2}{13},\cdots$ values,
that is, $\alpha=2/(5+4n),$ $n=0,1,2,\ldots$, so that the pole lies inside
the contour. We now write Equation (23) as%
\begin{equation}
T(t)=T(0)\left[ \frac{e^{i\lambda_{n}^{1/\alpha}t}}{\alpha}%
-F_{\alpha}(\sigma;t)\right] ,\text{ }\sigma=\lambda_{n}i^{\alpha},
\end{equation}
where%
\begin{equation}
F_{\alpha}(\sigma;t)=\frac{\sigma\sin\alpha\pi}{\pi}\int_{0}^{\infty}\frac{%
e^{-xt}x^{\alpha-1}dx}{x^{2\alpha}-2\sigma\cos\alpha\pi\text{ }%
x^{\alpha}+\sigma^{2}}.
\end{equation}

In previous works, the above integral was assumed to be a decaying
exponential, hence neglected in subsequent calculations [18, 22-24]. This
resulted in probabilities greater than $1.$ We now show that an exact
treatment of this integral, which can not be evaluated analytically by
standard techniques, proves otherwise. Using the elegant language of Fox's
H- functions, details of which are presented in Appendices B and C, we
evaluate the above integral exactly and show that it has an oscillatory part
along with a Mittag-Leffler function as

\begin{equation}
F_{\alpha}(\sigma;t)=\pm\frac{e^{i\lambda_{n}^{1/\alpha}t}}{\alpha}%
-E_{\alpha }(\lambda_{n}i^{\alpha}t^{\alpha}).
\end{equation}
Thus, the second method gives the time dependent part of the wave function
as 
\begin{equation}
T(t)=T(0)\left[ \frac{e^{i\lambda_{n}^{1/\alpha}t}}{\alpha}\mp\frac {%
e^{i\lambda_{n}^{1/\alpha}t}}{\alpha}+E_{\alpha}(\lambda_{n}i^{\alpha
}t^{\alpha})\right] .
\end{equation}
To be consistent with the robust result of Method I [Eq. (21)], we pick the
minus sign, hence the time dependent part of the wave function is again
obtained as 
\begin{equation}
T(t)=E_{\alpha}(\lambda_{n}i^{\alpha}t^{\alpha}),
\end{equation}
where without loss of any generality we have set $T(0)=1.$

\subsection{Probability and Asymptotic Behavior}

\subsubsection{Total Probability:}

Since the general time dependence of the separable solutions of the time
fractional Schr\"{o}dinger equation is established as the Mittag-Leffler
function with an imaginary argument [Eq. (28)], we can write the series 
\begin{align}
T(t) & =E_{\alpha}(\lambda_{n}i^{\alpha}t^{\alpha})=\sum_{\nu=0}^{\infty }%
\frac{(\lambda_{n}i^{\alpha}t^{\alpha})^{\nu}}{\Gamma(1+\alpha\nu)},\text{ }%
\left\vert \lambda_{n}i^{\alpha}t^{\alpha}\right\vert <1 \\
& =\left[ 1+\frac{\lambda_{n}i^{\alpha}t^{\alpha}}{\Gamma(1+\alpha)}+\frac{%
\left( \lambda_{n}i^{\alpha}t^{\alpha}\right) ^{2}}{\Gamma (1+2\alpha)}+%
\frac{\left( \lambda_{n}i^{\alpha}t^{\alpha}\right) ^{3}}{\Gamma(1+3\alpha)}%
+\cdots\right] ,
\end{align}
which can be separated into its real and imaginary parts as 
\begin{align}
E_{\alpha}(\lambda_{n}i^{\alpha}t^{\alpha}) & =E_{\alpha}^{R}(t)+iE_{\alpha
}^{I}(t),\text{ }0<\alpha<1, \\
& =\sum_{\nu=0}^{\infty}\frac{\lambda_{n}^{\nu}\left[ \cos\frac{\nu\alpha \pi%
}{2}\right] t^{\alpha\nu}}{\Gamma(1+\alpha\nu)}+i\sum_{\nu=0}^{\infty }\frac{%
\lambda_{n}^{\nu}\left[ \sin\frac{\nu\alpha\pi}{2}\right] t^{\alpha\nu}}{%
\Gamma(1+\alpha\nu)}.
\end{align}
When $\alpha=1,$ naturally, $E_{\alpha}(\lambda_{n}i^{\alpha}t^{\alpha})$
becomes the Euler equation, that is, 
\begin{align}
T_{\alpha=1}(t) & =\cos\lambda_{n}t+i\sin\lambda_{n}t \\
& =e^{i\lambda_{n}t}.
\end{align}
The eigenvalues, $\lambda_{n},$ come from the solution of the space part of
the Schr\"{o}dinger equation with the appropriate boundary conditions [Eq.
(11)]. Note that we can also write Equation (32) as 
\begin{equation}
T(t)=E_{\alpha}\left( \lambda_{n}\cos^{1/\nu}\left( \frac{\nu\alpha\pi}{2}%
\right) \text{ }t^{\alpha}\right) +iE_{\alpha}\left(
\lambda_{n}\sin^{1/\nu}\left( \frac{\nu\alpha\pi}{2}\right) \text{ }%
t^{\alpha}\right) ,
\end{equation}
which in terms of $H-$functions becomes (Appendix B, [25]) 
\begin{align}
T(t) & =\left[ H_{1,2}^{1,1}\left( \left. -\lambda_{n}\cos^{1/\nu}\left( 
\frac{\nu\alpha\pi}{2}\right) \text{ }t^{\alpha}\right\vert
_{(0,1),(0,\alpha )}^{(0,1)}\right) \right.  \notag \\
& \left. +iH_{1,2}^{1,1}\left( \left. -\lambda_{n}\sin^{1/\nu}\left( \frac{%
\nu\alpha\pi}{2}\right) \text{ }t^{\alpha}\right\vert _{(0,1),(0,\alpha
)}^{(0,1)}\right) \right] .
\end{align}

An important consequence of this result is that for the time fractional Schr%
\"{o}dinger equation, the total probability, $\int_{-\infty}^{+\infty
}\left\vert \Psi(x,t)\right\vert dx,$ which is a function of time: 
\begin{align}
\int_{-\infty}^{+\infty}\left\vert \Psi(x,t)\right\vert dx & =\left\vert
T(t)\right\vert ^{2}\int_{-\infty}^{+\infty}\left\vert X(x)\right\vert ^{2}dx
\notag \\
& =\left[ \left\vert E_{\alpha}^{R}(t)\right\vert ^{2}+\left\vert E_{\alpha
}^{I}(t)\right\vert ^{2}\right] \int_{-\infty}^{+\infty}\left\vert
X(x)\right\vert dx  \notag \\
& =\left\vert E_{\alpha}^{R}(t)\right\vert ^{2}+\left\vert
E_{\alpha}^{I}(t)\right\vert ^{2},
\end{align}
is not conserved. The normalization constant is fixed by the total
probability at $t=0$: 
\begin{equation}
\int_{-\infty}^{+\infty}\left\vert \Psi(x,0)\right\vert dx=\int_{-\infty
}^{+\infty}\left\vert X(x)\right\vert dx=1.
\end{equation}

\subsubsection{Small Time Behavior:}

For small times, we use the $H-$function representation of the
Mittag-Leffler function: 
\begin{equation}
E_{\alpha }(z)=H_{1,2}^{1,1}\left( -z_{(0,1),(0,\alpha )}^{(0,1)}\right) ,
\end{equation}%
and write the time dependence as 
\begin{equation}
T(t)=H_{1,2}^{1,1}\left( \left. -\lambda _{n}i^{\alpha }t^{\alpha
}\right\vert _{(0,1),(0,\alpha )}^{(0,1)}\right) .
\end{equation}%
Using Equation (C3) we again obtain the series [Eq. (32)] 
\begin{align}
T(t)& =\left[ 1+\frac{\cos (\alpha \pi /2)}{\Gamma (1+\alpha )}(\lambda
_{n}t^{\alpha })+\frac{\cos (\alpha \pi )}{\Gamma (1+2\alpha )}(\lambda
_{n}t^{\alpha })^{2}+\frac{\cos (3\alpha \pi /2)}{\Gamma (1+3\alpha )}%
(\lambda _{n}t^{\alpha })^{3}+\cdots \right]  \notag \\
& +i\left[ \frac{\sin (\alpha \pi /2)}{\Gamma (1+\alpha )}(\lambda
_{n}t^{\alpha })+\frac{\sin (\alpha \pi )}{\Gamma (1+2\alpha )}(\lambda
_{n}t^{\alpha })^{2}+\frac{\sin (3\alpha \pi /2)}{\Gamma (1+3\alpha )}%
(\lambda _{n}t^{\alpha })^{3}+\cdots \right] ,
\end{align}%
which for small times gives the total probability as

\begin{align}
\int_{-\infty }^{+\infty }\left\vert \Psi (x,t)\right\vert ^{2}dx&
=\left\vert T(t)\right\vert ^{2}\int_{-\infty }^{+\infty }\left\vert
X(x)\right\vert ^{2}dx \\
& \simeq \left( 1+\frac{2\cos (\alpha \pi /2)(\lambda _{n}t^{\alpha })}{%
\alpha \Gamma (\alpha )}+0(t^{2\alpha })\right) .
\end{align}%
For $\lambda _{n}<0,$ this makes the total probability $\leq 1$ for small
times.

\subsubsection{Large Time Behavior:}

For large times, we use the expansion in Equation (C7) to write 
\begin{align}
T(t)& =\sum_{v=0}^{\infty }\frac{-1}{\Gamma (1-\nu \alpha )(\lambda
_{n}i^{\alpha }t^{\alpha })^{1+\nu }} \\
& =\left[ -\frac{1}{\lambda _{n}i^{\alpha }t^{\alpha }}-\frac{1}{\Gamma
(1-\alpha )(\lambda _{n}i^{\alpha }t^{\alpha })^{2}}-\frac{1}{\Gamma
(1-2\alpha )(\lambda _{n}i^{\alpha }t^{\alpha })^{3}}-\cdots \right] .
\end{align}%
We can also write this in terms of its real and imaginary parts as 
\begin{align}
T(t)& =\left[ -\frac{\cos (\alpha \pi /2)}{\lambda _{n}t^{\alpha }}-\frac{%
\cos (\alpha \pi )}{\Gamma (1-\alpha )(\lambda _{n}t^{\alpha })^{2}}-\frac{%
\cos (3\alpha \pi /2)}{\Gamma (1-2\alpha )(\lambda _{n}t^{\alpha })^{3}}%
+\cdots \right]  \notag \\
& +i\left[ \frac{\sin (\alpha \pi /2)}{\lambda _{n}t^{\alpha }}+\frac{\sin
(\alpha \pi )}{\Gamma (1-\alpha )(\lambda _{n}t^{\alpha })^{2}}+\frac{\sin
(3\alpha \pi /2)}{\Gamma (1-2\alpha )(\lambda _{n}t^{\alpha })^{3}}+\cdots %
\right] ,
\end{align}%
In agreement with Diethelm et. al. [28], for large times, the total
probability decays as 
\begin{equation}
\left\vert T(t)\right\vert ^{2}\varpropto \frac{1}{\lambda
_{n}^{2}t^{2\alpha }}.
\end{equation}

Since the wave function is zero on and outside the boundary, the energy does
not leak out. Due to the presence of the fractional time derivative, the
particle inside the infinite well is not free according to the ordinary Schr%
\"{o}dinger equation. In Section V, we introduce the \textit{effective
potential} in terms of the ordinary Schr\"{o}dinger equation, where the
imaginary part of the complex effective potential describes the dissipation
process. Hence, the energy is basically exchanged within the box, between
the physical agent that generates the effective potential and the particle.

\section{The Box Problem}

\ \ We now consider a particle in an infinite potential well, where 
\begin{equation}
V(x)=\left\{ 
\begin{tabular}{lll}
$0$ & , & $0<x<a$ \\ 
&  &  \\ 
$\infty $ & , & elsewhere%
\end{tabular}%
\ \ \ \ \ \ \ \ \ \ \ \ \ \ \ \ \right. .
\end{equation}%
For the separable solutions of the time fractional schr\"{o}dinger equation
[Eq. (10)], we have already determined the time dependent part [Eq. (28)] of
the wave function, hence we can write $\Psi (x,t)=E_{\alpha }(\lambda
_{n}i^{\alpha }t^{\alpha })X(x),$ where the spatial part comes from the
solution of Equation (11) with the appropriate boundary conditions. For\ the
particle in a box with impenetrable walls, we solve 
\begin{equation}
D_{\alpha }\frac{d^{2}X(x)}{dx^{2}}=\lambda _{n}X(x),\text{ \ }\Psi
(0,t)=\Psi (a,t)=0,
\end{equation}%
which yields the eigenvalues and the eigenfunctions as 
\begin{equation}
\lambda _{n}=-D_{\alpha }\left( \frac{n\pi }{a}\right) ^{2},\text{ \ }%
X_{n}(x)=C_{0}\sin \left( \frac{n\pi }{a}x\right) ,\text{ }n=1,2,\ldots ,%
\text{ }D_{\alpha }>0.
\end{equation}%
Since we are considering the time fractional Schr\"{o}dinger equation, non
locality is in terms of time, which means that the system has memory. For
the separable solutions, the differential equation to be solved for the
space part of the wave equation is an ordinary differential equation, hence
the boundary conditions used in Eq. (49) are the natural boundary conditions
for impenetrable walls. Using the normalization condition $\int \left\vert
\Psi _{n}(x,0)\right\vert ^{2}dx=1,$ we write the complete wave function as 
\begin{equation}
\Psi _{n}(x,t)=\sqrt{\frac{2}{a}}\left[ \sin \left( \frac{n\pi }{a}x\right) %
\right] E_{\alpha }(\lambda _{n}i^{\alpha }t^{\alpha }).
\end{equation}%
The total probability is time dependent, which for small times behaves as 
\begin{align}
\int \left\vert \Psi _{n}(x,t)\right\vert ^{2}dx& =\left\vert E_{\alpha
}^{R}(t)\right\vert ^{2}+\left\vert E_{\alpha }^{I}(t)\right\vert ^{2} \\
& \simeq 1-\frac{2\cos (\alpha \pi /2)(\left\vert \lambda _{n}\right\vert
t^{\alpha })}{\alpha \Gamma (\alpha )}+0(t^{2\alpha }),
\end{align}%
and for large times decays as 
\begin{equation}
\lim_{t\rightarrow \infty }\int \left\vert \Psi _{n}(x,t)\right\vert
^{2}dx\thicksim \frac{1}{\lambda _{n}^{2}t^{2\alpha }}\rightarrow 0.
\end{equation}%
This is in contrast with the previous works [18, 22-24], where for large
times the total probability is constant and greater than 1: 
\begin{equation}
\lim_{t\rightarrow \infty }\int \left\vert \Psi _{n}(x,t)\right\vert ^{2}dx=%
\frac{1}{\alpha ^{2}}\geq 1.
\end{equation}

\subsection{The Energy Operator and the Box Problem}

For the time fractional Schr\"{o}dinger equation, we take the new energy
operator as 
\begin{equation}
E=-\frac{\hbar}{i^{\alpha}}\frac{\partial^{\alpha}}{\partial t^{\alpha}},
\end{equation}
where the Hamiltonian is 
\begin{equation}
H_{\alpha}=-D_{\alpha}\hbar\frac{\partial^{2}}{\partial x^{2}}+V(x).
\end{equation}
With this operator, the energy values are real but time dependent:%
\begin{align}
E & =-\frac{\hbar}{i^{\alpha}}\int\Psi^{\ast}(x,t)\frac{\partial^{\alpha}}{%
\partial t^{\alpha}}\Psi(x,t)dx  \notag \\
& =-\hbar\lambda_{n}\left\vert E_{\alpha}(\lambda_{n}i^{\alpha}t^{\alpha
})\right\vert ^{2}\int\ \left\vert X(x)\right\vert ^{2}dx  \notag \\
& =-\hbar\lambda_{n}\left\vert E_{\alpha}(\lambda_{n}i^{\alpha}t^{\alpha
})\right\vert ^{2}
\end{align}

For the box problem, the energy eigenvalues are now obtained as 
\begin{eqnarray}
E_{n} &=&-\frac{\hbar }{i^{\alpha }}\int \Psi _{n}^{\ast }(x,t)\frac{%
\partial ^{\alpha }}{\partial t^{\alpha }}\Psi _{n}(x,t)dx, \\
&=&\frac{\hbar \pi ^{2}n^{2}D_{\alpha }}{a^{2}}\ \left[ \left\vert E_{\alpha
}^{R}(t)\right\vert ^{2}+\left\vert E_{\alpha }^{I}(t)\right\vert ^{2}\right]
\end{eqnarray}%
which for small times are given as \ 
\begin{equation}
E_{n}\simeq \frac{\hbar \pi ^{2}n^{2}D_{\alpha }}{a^{2}}\left( 1-\frac{2\cos
(\alpha \pi /2)(\left\vert \lambda _{n}\right\vert t^{\alpha })}{\alpha
\Gamma (\alpha )}\right) .  \notag
\end{equation}%
For large times $(\alpha \neq 1),$ the system dissipates all of its energy
as 
\begin{equation}
\lim_{t\rightarrow \infty }E_{n}\thicksim \frac{\hbar \pi ^{2}n^{2}D_{\alpha
}}{a^{2}\lambda _{n}^{2}t^{2\alpha }}\rightarrow 0.
\end{equation}%
In the limit as $\alpha \rightarrow 1,$ $D_{\alpha }\rightarrow $ $\hbar /2m$%
, the energy eigenvalues approach to their usual values predicted by the Schr%
\"{o}dinger equation: 
\begin{equation}
E_{n}=\frac{\hbar ^{2}\pi ^{2}n^{2}}{2ma^{2}}.
\end{equation}
Note that the ordinary energy operator, $i\hbar \dfrac{\partial }{\partial t}
$, yields complex eigenvalues, which only in the limit as $t\rightarrow
\infty $ approach to real values that are larger than their usual values in
Equation (62) by a factor of $1/\nu ^{2}$.

\section{The Effective Potential\textbf{\ }}

To gain a better understanding of the effects of the fractional time
derivative, we look for an effective potential in the (ordinary) Schr\"{o}%
dinger equation that yields the same wave function. For simplicity we will
set $V(x)=0$ in Equation (11). We now use the following useful formula [18]: 
\begin{equation}
\frac{dT(t)}{dt}=_{0}^{C}D_{t}^{1-\alpha }\left[ _{0}^{C}D_{t}^{\alpha }T(t)%
\right] +\frac{\left[ _{0}^{C}D_{t}^{\alpha }T(t)\right] _{t=0}}{\Gamma
(\alpha )t^{1-\alpha }},
\end{equation}%
which is particularly helpful in isolating and exploring the effects of
fractional time derivatives in terms of an effective potential. Using
separable solutions of the time fractional Schr\"{o}dinger equation [Eq.
(12)], we can write 
\begin{equation}
\frac{dT(t)}{dt}=i^{\alpha }\lambda _{n}\left[ _{0}^{C}D_{t}^{1-\alpha }T(t)+%
\frac{T(0)}{\Gamma (\alpha )t^{1-\alpha }}\right] .
\end{equation}%
The quantity inside the square brackets is nothing but the Riemann-Liouville
derivative [Eq. (A5)], thus 
\begin{equation}
\frac{dT(t)}{dt}=i^{\alpha }\lambda _{n}\left[ _{0}^{R-L}D_{t}^{1-\alpha
}T(t)\right] .
\end{equation}

To define an effective potential for the separable solutions, we also write
the (ordinary) Schr\"{o}dinger equation as%
\begin{equation}
i\hbar\frac{dT(t)}{dt}X(x)=-\frac{\hbar^{2}}{2m}\frac{d^{2}X(x)}{dx^{2}}%
T(t)+V_{eff.}(t)X(x)T(t),
\end{equation}
and substitute the wave function found from the solution of the time
fractional Schr\"{o}dinger equation [Eq. (65) and Eq. (11) with $V(x)=0$]: 
\begin{equation}
i\hbar i^{\alpha}\lambda_{n}\left[ _{0}^{R-L}D_{t}^{1-\alpha}T(t)\right]
X(x)=-\frac{\hbar^{2}}{2m}\frac{\lambda_{n}}{D_{\alpha}}%
X(x)T(t)+V_{eff.}(t)X(x)T(t),
\end{equation}
This yields the effective potential as 
\begin{equation}
V_{eff.}(t)=i^{1+\alpha}\hbar\lambda_{n}\frac{_{0}^{R-L}D_{t}^{1-\alpha}T(t)%
}{T(t)}+\frac{\hbar^{2}}{2m}\frac{\lambda_{n}}{D_{\alpha}}.
\end{equation}
Since the time dependence is given as $T(t)=E_{\alpha}(\lambda_{n}i^{\alpha
}t^{\alpha}),$ we can also write 
\begin{equation}
V_{eff.}(t)=i^{1+\alpha}\hbar\lambda_{n}\frac{_{0}^{R-L}D_{t}^{1-\alpha
}E_{\alpha}(\lambda_{n}i^{\alpha}t^{\alpha})}{E_{\alpha}(\lambda_{n}i^{%
\alpha }t^{\alpha})}+\frac{\hbar^{2}}{2m}\frac{\lambda_{n}}{D_{\alpha}}.
\end{equation}

In the limit as $\alpha\rightarrow1,$ $D_{\alpha}\rightarrow\dfrac{\hbar}{2m}%
,$ the Mittag-Leffler function becomes the exponential function, $%
T(t)=e^{i\lambda_{n}t}$, hence the effective potential vanishes as it
should. With this effective potential, the Schr\"{o}dinger equation yields
the same wave function as the time fractional Schr\"{o}dinger equation and
satisfies the same boundary conditions. In other words, the wave function
found from the time fractional Schr\"{o}dinger equation, satisfies the Schr%
\"{o}dinger equation with the above effective potential.

Using the $H-$function [25] representation of the Mittag-Leffler function
(Appendix B): 
\begin{equation}
E_{\alpha }(\lambda _{n}i^{\alpha }t^{\alpha })=H_{1,2}^{1,1}\left( \left.
-\lambda _{n}i^{\alpha }t^{\alpha }\right\vert _{(0,1),(0,\alpha
)}^{(0,1)}\right)
\end{equation}%
and Equation (B7) for the Riemann-Liouville derivative of the $H-$function,
we can also write the following closed expression for the effective
potential: 
\begin{equation}
V_{eff.}(t)=\frac{\left( i^{1+\alpha }\hbar \lambda _{n}\right) }{%
t^{1-\alpha }}\frac{H_{2,3}^{1,2}\left( \left. -\lambda _{n}i^{\alpha
}t^{\alpha }\right\vert _{(0,1),(0,\alpha ),(1-\alpha ,\alpha )}^{(0,\alpha
),(0,1)}\right) }{H_{1,2}^{1,1}\left( \left. -\lambda _{n}i^{\alpha
}t^{\alpha }\right\vert _{(0,1),(0,\alpha )}^{(0,1)}\right) }+\frac{\hbar
^{2}}{2m}\frac{\lambda _{n}}{D_{\alpha }}.
\end{equation}%
Using the symmetries of the $H-$function [25], we can further simplify the
above equation as 
\begin{equation}
V_{eff.}(t)=\frac{\left( i^{1+\alpha }\hbar \lambda _{n}\right) }{%
t^{1-\alpha }}\frac{H_{1,2}^{1,1}\left( \left. \left( (-\lambda
_{n})^{1/\alpha }it\right) ^{\alpha }\right\vert _{(0,1),(1-\alpha ,\alpha
)}^{(0,1)}\right) }{H_{1,2}^{1,1}\left( \left. -\lambda _{n}i^{\alpha
}t^{\alpha }\right\vert _{(0,1),(0,\alpha )}^{(0,1)}\right) }+\frac{\hbar
^{2}}{2m}\frac{\lambda _{n}}{D_{\alpha }},
\end{equation}%
or as 
\begin{equation}
V_{eff.}(t)=\frac{\left( i^{1+\alpha }\hbar \lambda _{n}\right) }{%
t^{1-\alpha }}\frac{E_{\alpha ,\alpha }(\lambda _{n}i^{\alpha }t^{\alpha })}{%
E_{\alpha }(\lambda _{n}i^{\alpha }t^{\alpha })}+\frac{\hbar ^{2}}{2m}\frac{%
\lambda _{n}}{D_{\alpha }},
\end{equation}%
where 
\begin{equation}
E_{\alpha ,\beta }(z)=\sum_{k=0}^{\infty }\frac{z^{k}}{\Gamma (\alpha
k+\beta )},\text{ }\alpha ,\beta \in \mathbb{C},\text{ }\func{Re}(a)>0,\text{
}\func{Re}(\beta )>0,\text{ }z\in \mathbb{C},
\end{equation}%
is the generalized Mittag-Leffler function, and $E_{\alpha ,1}(z)=E_{\alpha
}(z)$ [25].

\subsection{Hamiltonian}

Since the effective potential is time dependent and complex, the energies
corresponding to the operator $i\hbar\dfrac{\partial}{\partial t}$ are also
time dependent and complex, that is, the ordinary Hamiltonian operator: 
\begin{equation}
H=-\frac{\hbar^{2}}{2m}\frac{d^{2}}{dx^{2}}+V_{eff.}(t)
\end{equation}
is not hermitian. However, complex potentials in quantum mechanics have
found interesting applications in the study of dissipative systems [29-32].

Let us now write the effective potential for the time fractional Schr\"{o}%
dinger equation explicitly. Using Equation (73) and the expression 
\begin{equation}
E_{\alpha ,\alpha }(\lambda _{n}i^{\alpha }t^{\alpha })=\sum_{k=0}^{\infty }%
\frac{\lambda _{n}^{k}\cos (k\alpha \pi /2)t^{\alpha k}}{\Gamma (1+k\alpha )}%
+i\sum_{k=0}^{\infty }\frac{\lambda _{n}^{k}\sin (k\alpha \pi /2)t^{\alpha k}%
}{\Gamma (1+k\alpha )},
\end{equation}%
we can write the effective potential in terms of its real and imaginary
parts as 
\begin{equation}
V_{eff.}(t)=V_{eff.}^{R}(t)+iV_{eff.}^{I}(t),
\end{equation}%
where 
\begin{align}
V_{eff.}^{R}(t)& =\frac{\hbar ^{2}\lambda _{n}}{2mD_{\alpha }}-\frac{\hbar
\lambda _{n}\sin (\alpha \pi /2)t^{\alpha -1}}{\Gamma (\alpha )}  \notag \\
& -\hbar \lambda _{n}^{2}\sin (\alpha \pi )\left( \frac{1}{\Gamma (2\alpha )}%
-\frac{1}{\Gamma (1+\alpha )\Gamma (\alpha )}\right) t^{2\alpha -1}+\cdots 
\end{align}%
and 
\begin{align}
V_{eff.}^{I}(t)& =\frac{\hbar ^{2}\lambda _{n}}{2mD_{\alpha }}-\frac{\hbar
\lambda _{n}\cos (\alpha \pi /2)t^{\alpha -1}}{\Gamma (\alpha )}  \notag \\
& +\hbar \lambda _{n}^{2}\sin (\alpha \pi )\left( \frac{1}{\Gamma (1+\alpha
)\Gamma (\alpha )}+\frac{\cos (\alpha \pi )}{\Gamma (2\alpha )}-\frac{2\cos
^{2}(\alpha \pi /2)}{\Gamma (\alpha )\Gamma (1+\alpha )}\right) t^{2\alpha
-1}+\cdots \text{ }.
\end{align}%
The imaginary part of the complex effective potential plays an important
role in describing the decay of the energy of the stationary states of the
time-fractional Schr\"{o}dinger equation.

\section{The Mittag-Leffler Function, $E_{\protect\alpha }(\protect\lambda %
_{n}i^{\protect\alpha }t^{\protect\alpha }),$ and it's Purely Decaying and
Oscillating Parts}

Solving\ the Schr\"{o}dinger equation [Eq. (66)] with the above effective
potential, we now obtain the following expression for the time dependent
part of the separable solutions of the time fractional Schr\"{o}dinger
equation:

\begin{equation}
T(t)=\left[ \exp \left( \dfrac{1}{\hbar }\dint_{0}^{t}V_{eff.}^{I}(t^{\prime
})dt^{\prime }\right) \right] \exp \left( \dfrac{i}{\hbar }\left[ \dfrac{%
\hbar ^{2}\lambda _{n}t}{2mD_{\alpha }}-\dint_{0}^{t}V_{eff.}^{R}(t^{\prime
})dt^{\prime }\right] \right) ,
\end{equation}%
where $\lambda _{n}$ is a separation constant. This is naturally equivalent
to the solution found before [Eq. (28)], that is, $T(t)=E_{\alpha }(\lambda
_{n}i^{\alpha }t^{\alpha })$. Note that when $\alpha =1,$ the effective
potential vanishes, hence the time dependence reduces to%
\begin{equation}
T(t)=\exp \left( -\frac{iE_{n}t}{\hbar }\right) ,\text{ }E_{n}=\dfrac{\hbar
^{2}\lambda _{n}}{2mD_{1}}.
\end{equation}

In other words, the time dependent part of the separable solution of the
time fractional Schr\"{o}dinger equation, $E_{\alpha }(\lambda _{n}i^{\alpha
}t^{\alpha }),$ can be written, not as the sum, but as the product of its
purely oscillating and purely exponentially decaying/growing parts. Despite
the fact that the effective potential corresponding to the time fractional
Schr\"{o}dinger equation is complex, the advantage of the effective
potential approach is that one can use the mathematical structure of the Schr%
\"{o}dinger theory.

Finally, using the expansion 
\begin{align}
V_{eff.}(t)& \simeq \left( \frac{\hbar ^{2}\lambda _{n}}{2mD_{\alpha }}-%
\frac{\hbar \lambda _{n}\sin (\alpha \pi /2)}{\Gamma (\alpha )}t^{\alpha
-1}+0(t^{2\alpha -1})\right)  \notag \\
& +i\left( \frac{\hbar \lambda _{n}\cos (\alpha \pi /2)}{\Gamma (\alpha )}%
t^{\alpha -1}+0(t^{2\alpha -1})\right) ,
\end{align}%
the small time behavior of $T(t),$ to lowest order, becomes 
\begin{equation}
T(t)\simeq \left[ \exp \left( \frac{\lambda _{n}\cos (\alpha \pi /2)}{\alpha
\Gamma (\alpha )}t^{\alpha }\right) \right] .\left[ \exp \left( \frac{i}{%
\hbar }\left( \frac{\hbar \lambda _{n}\sin (\alpha \pi /2)}{\alpha \Gamma
(\alpha )}t^{\alpha }\right) \right) \right] .
\end{equation}%
For the box problem, where $\lambda _{n}=-D_{\alpha }\left( \frac{n\pi }{a}%
\right) ^{2},$ $n=1,2,\ldots ,$ $D_{\alpha }>0,$ the total probability
decays exponentially, which for small times becomes 
\begin{equation}
\left\vert T(t)\right\vert ^{2}\simeq 1-\frac{2\cos (\alpha \pi /2)}{\alpha
\Gamma (\alpha )}\left\vert \lambda _{n}\right\vert t^{\alpha }+0(t^{2\alpha
}).
\end{equation}%
This is in accordance with our previous result [Eq. (43)] obtained from $%
T(t)=E_{\alpha }(\lambda _{n}i^{\alpha }t^{\alpha })$.

\section{Conclusions}

The fact that anomalous diffusion can be studied by fractional calculus has
attracted researchers from many different branches of science and
engineering into this intriguing branch of mathematics [1-6]. Since 60's,
successful examples are given in random walk, anomalous diffusion, economics
and finance. Introduction of fractional calculus into a certain branch of
science is usually initiated by replacing certain derivatives in the
evolution or the transport equations with their fractional counterparts. A
general feature of these applications is that replacing a time derivative
with its fractional counterpart basically introduces memory effects into the
system and makes the process non Markovian, while a replacement of a space
derivative introduces global or non local effects. However, it is important
to note that not all such effects can be covered by fractional
generalizations of the basic equations [20, 21].

In this paper, we first established the time dependent part of the separable
solutions of the time fractional Schr\"{o}dinger equation as the
Mittag-Leffler function with an imaginary argument by two different methods.
This showed that the total probability is $\leq 1$ and decays with time. We
also introduced the effective potential approach, where with the effective
potential, the (ordinary) Schr\"{o}dinger equation yields the same wave
function, with the same boundary conditions, as that the time fractional Schr%
\"{o}dinger equation.

For the energy, the standard operator used in Schr\"{o}dinger theory, $%
E=i\hbar \dfrac{\partial }{\partial t},$ yields complex values since the
corresponding Hamiltonian is complex and non hermitian. \ On the other hand,
the operator $E=-\dfrac{\hbar }{i^{\alpha }}\dfrac{\partial ^{\alpha }}{%
\partial t^{\alpha }},$ which follows naturally from the time fractional Schr%
\"{o}dinger equation, yields real but time dependent energies [Eq. (58)]. In
the light of these, we also discussed the particle in a box problem, which
is essentially the prototype of a device or a detector with internal degrees
of freedom. With the operator $E=-\dfrac{\hbar }{i^{\alpha }}\dfrac{\partial
^{\alpha }}{\partial t^{\alpha }},$ the energy eigenvalues are real and time
dependent, and decay with the imaginary part of the complex effective
potential [Eq. (79)] as 
\begin{equation}
E_{n}=\frac{\hbar \pi ^{2}D_{\alpha }n^{2}}{a^{2}}\exp \left( \dfrac{2}{%
\hbar }\dint_{0}^{t}V_{eff.}^{I}(t^{\prime })dt^{\prime }\right) .
\end{equation}%
Since the wave function is zero on and outside the boundary, the energy is
exchanged within the box between the source of the effective potential and
the particle. The imaginary part of the complex potential describes the
dissipative processes involved.

Complex potentials have been very useful in the study of dissipative
processes in both classical and quantum physics. Sinha et. al. [29] have
extended the factorization technique of Kuru and Negro [30] to study complex
potentials in classical systems that are analogues of non hermitian quantum
mechanical systems. In the context of quantum device modeling, a general non
hermitian Hamiltonian operator has been investigated by Ferry et. al. [31].
They have also discussed the general behavior of complex potentials with
applications to ballistic quantum dots and its implications for trajectories
and histories in the dots. Barraf [32] have used uniform complex potentials
to model particle capture from an incident beam by quantum wells. We have
pointed out that there is a certain ambiguity in writing the time fractional
Schr\"{o}dinger equation. For separable solutions, the time fractional Schr%
\"{o}dinger equation that Naber [18] and Dong and Xu [22] used gives the
time dependence of the wave function as $E_{\alpha }(\lambda
_{n}(-i)^{\alpha }t^{\alpha }).$ However, for the total probability [Eq.
(37)] and the energy [Eq. (58)], both conventions yield the same time
dependence since 
\begin{equation}
\left\vert E_{\alpha }(\lambda _{n}(-i)^{\alpha }t^{\alpha })\right\vert
=\left\vert E_{\alpha }(\lambda _{n}i^{\alpha }t^{\alpha })\right\vert ,%
\text{ }0<\alpha <1.
\end{equation}

\appendix

\section{Basic Definitions}

\textbf{Riemann-Liouville Definition of Differintegral:}

The basic definition of fractional derivative and integral, that is,
differintegral, is the Riemann-Liouville (R-L) definition:

For $\ q<0,$ the R-L fractional integral is evaluated by using the formula 
\begin{equation}
\text{\ }\left[ \frac{d^{q}f}{[d(t-a)]^{q}}\right] =\frac{1}{\Gamma(-q)}%
\int_{a}^{t}[t-t^{\prime}]^{-q-1}f(t^{\prime})dt^{\prime},\text{ \ \ \ \ }\
q<0.
\end{equation}

For fractional derivatives, $q\geq0,$ the above integral is divergent, hence
the R-L formula is modified as [6] 
\begin{equation}
\left[ \frac{d^{q}f}{[d(t-a)]^{q}}\right] =\frac{d^{n}}{dt^{n}}\left[ \frac{1%
}{\Gamma(n-q)}\int_{a}^{t}[t-t^{\prime}]^{-(q-n)-1}f(t^{\prime })dt^{\prime}%
\right] ,\text{\ \ \ \ }q\geq0,\text{ }n>q,
\end{equation}
where the integer $n$ must be chosen as the smallest integer satisfying\ $%
(q-n)<0$.

For $0<q<1$ and $a=0,$ the Riemann-Liouville fractional derivative becomes 
\begin{equation}
\left[ \frac{d^{q}f(t)}{dt^{q}}\right] _{R-L}=\frac{1}{\Gamma (1-q)}\frac{d}{%
dx}\int_{0}^{t}\frac{f(t^{\prime })d\tau }{(t-t^{\prime })^{q}},\text{ }%
0<q<1.
\end{equation}

\textbf{Caputo Fractional Derivative:}

In 60's Caputo introduced a new definition of fractional derivative [1-6]:%
\begin{equation}
\left[ \frac{d^{q}f(t)}{dt^{q}}\right] _{C}=\frac{1}{\Gamma(1-q)}\int
_{0}^{t}\left( \frac{df(\tau)}{d\tau}\right) \frac{d\tau}{(t-\tau)^{q}},%
\text{ }0<q<1,
\end{equation}
which was used by him to model dissipation effects in linear
viscosity.\bigskip\ The two derivatives are related by%
\begin{equation}
\left[ \frac{d^{q}f(t)}{dt^{q}}\right] _{C}=\left[ \frac{d^{q}f(t)}{dt^{q}}%
\right] _{R-L}-\frac{t^{-q}f(0)}{\Gamma(1-q)},\text{ }0<q<1.
\end{equation}

Laplace transforms of the Riemann-Liouville and the Caputo derivative are
given as (Supplements of Ch. 14 of [6]) 
\begin{align}
\pounds \left\{ _{0}^{R-L}\mathbf{D}_{t}^{q}f(t)\right\} & =s^{q}\widetilde{f%
}(s)-\sum_{k=0}^{n-1}s^{k}\left. \left( _{0}^{R-L}\mathbf{D}%
_{t}^{q-k-1}f(t)\right) \right\vert _{t=0},\text{ }n-1<q\leq n, \\
\pounds \left\{ _{0}^{C}\mathbf{D}_{t}^{q}f(t)\right\} & =s^{q}\widetilde{f}%
(s)-\sum_{k=0}^{n-1}s^{q-k-1}\left. \frac{d^{k}f(t)}{dt^{k}}\right\vert
_{t=0},\text{ }n-1<q\leq n,
\end{align}
where $_{a}\mathbf{D}_{t}^{q}f(t)\equiv\dfrac{d^{q}f}{[d(t-a)]^{q}},$ and
wherever there is need for distinction we use the abbreviation `R-L' or `C'.
Since the Caputo derivative allows us to impose boundary conditions in terms
of ordinary derivatives, it has found widespread use.

\textbf{Riesz Derivative:}

In writing the space fractional diffusion equation or the space fractional
Schr\"{o}dinger equation, we use the Riesz derivative, which is defined with
respect to its Fourier transform (Supplements of Ch. 14 of [6]) 
\begin{equation}
\mathcal{F}\left\{ \mathbf{R}_{t}^{q}f(t)\right\} =-\left\vert
\omega\right\vert ^{q}g(\omega),\text{ }0<q<2,
\end{equation}
as 
\begin{equation}
\mathbf{R}_{t}^{q}f(t)=-\frac{1}{2\pi}\int_{-\infty}^{+\infty}\left\vert
\omega\right\vert ^{q}g(\omega)e^{i\omega t}d\omega,
\end{equation}
where $g(\omega)$ is the Fourier transform of $f(t)$. Note that 
\begin{equation}
\mathbf{R}_{t}^{2}f(t)=\frac{d^{2}}{dt^{2}}f(t).
\end{equation}

\section{Fox's H-Function}

In 1961 Fox introduced the $H-$function, which gives a general way of
expressing a wide range of functions encountered in applied mathematics. $H-$
function provides an elegant and an efficient formalism to handle problems
in fractional calculus. Fox's $H-$ function is a generalization of the
Meijer's $G-$function and is defined with respect to a Mellin-Barnes type
integral [25-27]: 
\begin{align}
H_{p,q}^{m,n}(z) & =H_{p,q}^{m,n}\left( \left. z\right\vert
_{(b_{q},B_{q})}^{(a_{p},A_{p})}\right) =H_{p,q}^{m,n}\left( \left.
z\right\vert
_{(b_{1},B_{1}),\ldots,(b_{q},B_{q})}^{(a_{1},A_{1}),\ldots,(a_{p},A_{p})}%
\right) \\
& =\frac{1}{2\pi i}\int_{C}h(s)z^{-s}ds,
\end{align}
where 
\begin{equation}
h(s)=\frac{\prod_{j=1}^{m}\Gamma(b_{j}+B_{j}s)\prod_{j=1}^{n}\Gamma
(1-a_{j}-A_{j}s)\ }{\prod_{j=n+1}^{p}\Gamma(a_{j}+A_{j}s)\prod_{j=m+1}^{q}%
\Gamma(1-b_{j}-B_{j}s)\ },
\end{equation}
$m,n,p,q$ are positive integers satisfying $0\leq n\leq p,$ $1\leq m\leq q,$
and empty products are taken as unity. Also, $A_{j},$ $j=1,\ldots,p,$ and $%
B_{j},$ $j=1,\ldots,q,$ are real positive numbers, and $a_{j},$ $%
j=1,\ldots,p,$ and $b_{j},$ $j=1,\ldots,q,$ are in general complex numbers
satisfying 
\begin{equation}
A_{j}(b_{h}+\nu)\neq B_{h}(a_{j}-\lambda-1)\text{ for }\nu,\lambda
=0,1,\ldots;\text{ }h=1,\ldots,m,\text{ }j=1,\ldots,n.
\end{equation}
The contour $C$ is such that the poles of $\Gamma(b_{j}+B_{j}s),$ $%
j=1,\ldots,m,$ are separated from the poles of $\Gamma(1-a_{j}-A_{j}s),$ $%
j=1,\ldots,n,$ such that the poles of $\Gamma(b_{j}+B_{j}s)$ lie to the left
of $C$, while the poles of $\Gamma(1-a_{j}-A_{j}s)$ are to the right of $C.$
The poles of the integrand are assumed to be simple. The $H-$function is
analytic for every $\left\vert z\right\vert \neq0$ when $\mu>0,$ and
analytic for $0<\left\vert z\right\vert <1/\beta$ when $\mu=0$, where $\mu$
and $\beta$ are defined as 
\begin{align}
\mu & =\sum_{j=1}^{q}B_{j}-\sum_{j=1}^{p}A_{j}, \\
\text{ }\beta & =\prod_{j=1}^{p}A_{j}^{A_{j}}\prod_{j=1}^{q}B_{j}^{-B_{j}}.\ 
\end{align}

Fox's $H-$function is very useful in the study of stochastic processes and
in solving fractional diffusion equations [33, 34]. For example, the
following useful formula for the Riemann-Liouville fractional derivative of
the $H-$function [33]: 
\begin{equation}
_{0}^{R-L}\mathbf{D}_{z}^{\beta}\left[ z^{a}H_{p,q}^{m,n}\left( \left.
(cz)^{b}\right\vert _{(b_{j},B_{j})}^{(a_{j},A_{j})}\right) \right]
=z^{a-\beta}H_{p+1,q+1}^{m,n+1}\left( \left. (cz)^{b}\right\vert
_{(b_{j},B_{j}),(\beta-a,b)}^{(-a,b),(a_{j},A_{j})}\right) ,
\end{equation}
where $a,b>0$ and $a+b\min(b_{j}/B_{j})>-1,$ $1\leq j\leq m,$ can be used to
find solutions to the fractional diffusion equation by tuning the indices to
appropriate values. Similarly, the Laplace transform of the $H-$function can
be obtained by using the formula [25] 
\begin{equation}
\pounds \left\{ x^{\rho-1}H_{p,q+1}^{m,n}\left( \left. ax^{\sigma
}\right\vert _{(b_{q},B_{q}),(1-\rho,\sigma)}^{(a_{p},A_{p})}\right)
\right\} =s^{-\rho}H_{p,q}^{m,n}\left( \left. as^{-\sigma}\right\vert
_{(b_{q},B_{q})}^{(a_{p},A_{p})}\right) ,
\end{equation}
where the inverse transform is given as 
\begin{equation}
\pounds ^{-1}\left\{ s^{-\rho}H_{p,q}^{m,n}\left( \left. as^{\sigma
}\right\vert _{(b_{q},B_{q})}^{(a_{p},A_{p})}\right) \right\} =x^{\rho
-1}H_{p+1,q}^{m,n}\left( \left. ax^{-\sigma}\right\vert
_{(b_{q},B_{q})}^{(a_{p},A_{p}),(\rho,\sigma)}\right) ,
\end{equation}
where 
\begin{equation}
\rho,\alpha,s\text{ }\in\mathbb{C},\text{ }\func{Re}(s)>0,\text{ }\sigma>0
\end{equation}
and 
\begin{equation}
\func{Re}(\rho)+\sigma\max_{1\leq i\leq n}\left[ \frac{1}{A_{i}}+\frac{\func{%
Re}(a_{i})}{A_{i}}\right] >0,\text{ }\left\vert \arg a\right\vert <\frac{%
\pi\theta}{2},\text{ }\theta=\alpha-\sigma.
\end{equation}
$\ $

\section{H-Function in Computable Form}

Given an $H-$function, we can compute, plot, and also study its asymptotic
forms [25-27] by using the following series expressions:

I) If the poles of 
\begin{equation}
\prod_{j=1}^{m}\Gamma(b_{j}+sB_{j})
\end{equation}
are simple, that is, if 
\begin{equation}
B_{h}(b_{j}+\lambda)\neq B_{j}(b_{h}+\nu),\text{ }j\neq h,\text{ }%
j,h=1,\ldots,m,\text{ }\lambda,\nu=0,1,2,\ldots,
\end{equation}
then the following expansion can be used: 
\begin{gather}
H_{p,p}^{m,n}(z)=\sum_{h=1}^{m}\sum_{\nu=0}^{\infty}  \notag \\
\frac{\left[ \prod_{j=1}^{\prime m}\Gamma(b_{j}-B_{j}(b_{h}+\nu )/B_{h})%
\right] \left[ \prod_{j=1}^{n}\Gamma(1-a_{j}+A_{j}(b_{h}+\nu )/B_{h})\right]
(-1)^{\nu}}{\left[ \prod_{j=m+1}^{q}\Gamma(1-b_{j}+B_{j}(b_{h}+\nu)/B_{h})%
\right] \left[ \prod_{j=n+1}^{p}\Gamma(a_{j}-A_{j}(b_{h}+\nu)/B_{h})\right] }
\notag \\
\times\frac{z^{(b_{h}+\nu)/B_{h}}}{\nu!B_{h}},
\end{gather}
where a prime in the product means $j\neq h.$ This series converges for all $%
z\neq0$ if $\mu>0,$ and for $0<\left\vert z\right\vert <1/\beta$, if $\mu=0,$
where $\mu$ and $\beta$ are defined as 
\begin{equation}
\mu=\sum_{j=1}^{q}B_{j}-\sum_{j=1}^{p}A_{j},\text{ }\beta=%
\prod_{j=1}^{p}A_{j}^{A_{j}}\prod_{j=1}^{q}B_{j}^{-B_{j}}.
\end{equation}

II) If the poles of 
\begin{equation}
\prod_{j=1}^{m}\Gamma(1-a_{j}-sA_{j})
\end{equation}
are simple: 
\begin{equation}
A_{h}(1-a_{j}+\nu)\neq A_{j}(1-a_{h}+\lambda),\text{ }j\neq h,\text{ }%
j,h=1,\ldots,n,\text{ }\lambda,\nu=0,1,2,\ldots,
\end{equation}
then the following expansion can be used: 
\begin{gather}
H_{p,p}^{m,n}(z)=\sum_{h=1}^{n}\sum_{\nu=0}^{\infty}  \notag \\
\frac{\left[ \prod_{j=1}^{\prime n}\Gamma(1-a_{j}-A_{j}(1-a_{h}+\nu )/A_{h})%
\right] \left[ \prod_{j=1}^{m}\Gamma(b_{j}+B_{j}(1-a_{h}+\nu )/A_{h})\right]
\ }{\left[ \prod_{j=m+1}^{q}\Gamma(1-b_{j}-B_{j}(1-a_{h}+\nu)/A_{h})\right] %
\left[ \prod_{j=n+1}^{p}\Gamma(a_{j}+A_{j}(1-a_{h}+\nu)/A_{h})\right] }, 
\notag \\
\times\frac{(-1)^{\nu}(\frac{1}{z})^{(1-a_{h}+\nu)/A_{h}}}{\nu!A_{h}},
\end{gather}
where a prime in the product means $j\neq h.$ This series converges for all $%
z\neq0$ if $\mu<0,$ and for $\left| z\right| >1/\beta$, if $\mu=0,$ where $%
\mu$ and $\beta$ are defined as above.

\section{The Mittag-Leffler Function}

To evaluate the Mittag-Leffler function we concentrate on the integral [Eq.
(25)] 
\begin{equation}
F_{\alpha }(\sigma ;t)=\frac{\sigma \sin \alpha \pi }{\pi }\int_{0}^{\infty }%
\frac{e^{-xt}x^{\alpha -1}dx}{x^{2\alpha }-2\sigma \cos \alpha \pi \text{ }%
x^{\alpha }+\sigma ^{2}}.
\end{equation}%
Since the Mittag-Leffler function represents behavior between a power law
and an exponential, it plays a crucial role in both scientific and
engineering applications. In this regard, it is imperative that we have a
closed expression for this integral, which in previous works [18, 22-24] was
assumed to be the purely decaying part of the Mittag-Leffler function with
an imaginary argument, hence neglected in subsequent calculation with
serious consequences. Since the standard methods do not allow us to evaluate
this integral, we are going to use of the language of $H-$functions.

We start by noting that the above integral requires the evaluation of the
Laplace transform 
\begin{equation}
\pounds \left\{ \frac{x^{\alpha-1}}{x^{2\alpha}-2\sigma\cos\alpha\pi\text{ }%
x^{\alpha}+\sigma^{2}}\right\} .
\end{equation}
We first express the function 
\begin{equation}
f(x)=\frac{1}{x^{2\alpha}-2\sigma\cos\alpha\pi\text{ }x^{\alpha}+\sigma^{2}},
\end{equation}
in terms of $H-$functions. For this, we evaluate its Mellin transform: 
\begin{equation}
\mathcal{M}\left\{ f(x)\right\} =\widehat{f}(s)=\int_{0}^{%
\infty}f(x)x^{s-1}dx.
\end{equation}
Using Equation (13.126) in Bayin [6], we evaluate the above integral to
obtain the Mellin transform 
\begin{equation}
\mathcal{M}\left\{ \frac{1}{x^{2}-2\sigma\cos\alpha\pi\text{ }x+\sigma^{2}}%
\right\} =-\frac{\pi(-\sigma)^{s-2}\sin\alpha(s-1)\pi}{\sin\pi s\text{ }%
\sin\alpha\pi},
\end{equation}
which after using the property 
\begin{equation}
\mathcal{M}\left\{ f(x^{\beta})\right\} =\frac{1}{\beta}\widehat{f}(\frac {s%
}{\beta}),\text{ }\beta>0,
\end{equation}
allows us to write 
\begin{equation}
\mathcal{M}\left\{ f(x)\right\} =-\frac{\pi(-\sigma)^{s/\alpha}}{%
\alpha\sigma^{2}}\frac{\sin(s-\alpha)\pi}{\sin\alpha\pi\text{ }\sin\frac {%
s\pi}{\alpha}}.
\end{equation}
We now use the relation [6] 
\begin{equation}
\sin\pi z=\frac{\pi}{\Gamma(z)\Gamma(1-z)},
\end{equation}
to write 
\begin{equation}
h(s)=\mathcal{M}\left\{ f(x)\right\} =\frac{\pi}{\alpha\sigma^{2}\sin
\alpha\pi}\frac{(-\sigma)^{s/\alpha}\Gamma(s/\alpha)\Gamma(1-s/\alpha)}{%
\Gamma(\alpha-s)\Gamma(1-\alpha-s)}.
\end{equation}
The inverse Mellin transform, which is defined as 
\begin{equation}
g(x)=\frac{1}{2\pi i}\int_{\gamma-i\infty}^{\gamma+i\infty}\widehat {g}%
(s)x^{-s}ds,
\end{equation}
where $\widehat{g}(s)$ is the Mellin transform of $g(x),$ allows us to write 
\begin{align}
\frac{1}{x^{2\alpha}-2\sigma\cos\alpha\pi\text{ }x^{\alpha}+\sigma^{2}} &
=\left( \frac{\pi}{\alpha\sigma^{2}\sin\alpha\pi}\right)  \notag \\
& \times\left[ \frac{1}{2\pi i}\int_{\gamma-i\infty}^{\gamma+i\infty}\frac{%
(-\sigma)^{s/\alpha}\Gamma(s/\alpha)\Gamma(1-s/\alpha)}{\Gamma
(\alpha-s)\Gamma(1-\alpha-s)}x^{-s}ds\right] .
\end{align}
Comparing with the definition of the $H-$function in Appendix (B): 
\begin{equation}
H_{p,q}^{m,n}\left( \left. z\right\vert
_{(b_{1},B_{1}),\ldots,(b_{q},B_{q})}^{(a_{1},A_{1}),\ldots,(a_{p},A_{p})}%
\right) =\frac{1}{2\pi i}\int_{C}h(s)z^{-s}ds,
\end{equation}
we read the values of the parameters as 
\begin{align}
m & =1,\text{ }n=1,\text{ }p=2,\text{ }q=2, \\
a_{1} & =0,\text{ }A_{1}=1/\alpha,\text{ }a_{2}=\alpha,\text{ }A_{2}=-1, \\
b_{1} & =0,\text{ }B_{1}=1/\alpha,\text{ }b_{2}=\alpha,\text{ }B_{2}=-1,
\end{align}
thus obtaining 
\begin{equation}
\frac{1}{x^{2\alpha}-2\sigma\cos\alpha\pi\text{ }x^{\alpha}+\sigma^{2}}%
=\left( \frac{\pi}{\alpha\sigma^{2}\sin\alpha\pi}\right) H_{2,2}^{1,1}\left(
\left. \frac{x}{(-\sigma)^{1/\alpha}}\right\vert _{(0,1/\alpha
),(\alpha,-1)}^{(0,1/\alpha),(\alpha,-1)}\right) .
\end{equation}

To evaluate $F_{\alpha}(\sigma;t),$ and finally the time dependence $T(t)$
of the wave function, we need the Laplace transform 
\begin{align}
\pounds \left\{ \frac{x^{\alpha-1}}{x^{2\alpha}-2\sigma\cos\alpha\pi\text{ }%
x^{\alpha}+\sigma^{2}}\right\} & =\left( \frac{\pi}{\alpha\sigma^{2}\sin%
\alpha\pi}\right)  \notag \\
& \times\pounds \left\{ x^{\alpha-1}H_{2,2}^{1,1}\left( \left. \frac {x}{%
(-\sigma)^{1/\alpha}}\right\vert
_{(0,1/\alpha),(\alpha,-1)}^{(0,1/\alpha),(\alpha,-1)}\right) \right\} .
\end{align}
Using the following expression of the Laplace transform [25]: 
\begin{equation}
\pounds \left\{ x^{\rho-1}H_{p,q}^{m,n}\left( \left. ax^{\sigma}\right\vert
_{(b_{q},B_{q})}^{(a_{p},A_{p})}\right) \right\}
=s^{-\rho}H_{p+1,q}^{m,n+1}\left( \left. as^{-\sigma}\right\vert
_{(b_{q},B_{q})}^{(1-\rho,\sigma),(a_{p},A_{p})}\right) ,
\end{equation}
with the replacements 
\begin{equation}
\rho\rightarrow\alpha,\text{ }a\rightarrow1/(-\sigma)^{1/\alpha},\text{ }%
\sigma\rightarrow1,\text{ }s\rightarrow t,
\end{equation}%
\begin{align}
m & =1,\text{ }n=1,\text{ }p=2,\text{ }q=2, \\
a_{1} & =0,\text{ }A_{1}=1/\alpha;\text{ }a_{2}=\alpha,\text{ }A_{2}=-1, \\
b_{1} & =0,\text{ }B_{1}=1/\alpha;\text{ }b_{2}=\alpha,\text{ }B_{2}=-1,
\end{align}
we obtain the needed transform as 
\begin{align}
\pounds \left\{ \frac{x^{\alpha-1}}{x^{2\alpha}-2\sigma\cos\alpha\pi\text{ }%
x^{\alpha}+\sigma^{2}}\right\} & =\frac{\pi}{\alpha\sigma^{2}\sin\alpha \pi}%
\frac{1}{t^{\alpha}}  \notag \\
& \times H_{2,3}^{2,1}\left( \left. (-\sigma)^{1/\alpha}t\right\vert
_{(\alpha,1),(1,1/\alpha),(1-\alpha,-1)}^{(0,1/\alpha),(1-\alpha,-1)}\right)
,
\end{align}
where we have used the relation [25] 
\begin{equation}
H_{p,q}^{m,n}\left( \left. z\right\vert
_{(b_{q},B_{q})}^{(a_{p},A_{p})}\right) =H_{q,p}^{n,m}\left( \left. \frac{1}{%
z}\right\vert _{(1-a_{p},A_{p})}^{(1-b_{q},B_{q})}\right) .
\end{equation}
Combining these results, we finally obtain the desired integral: 
\begin{equation}
F_{\alpha}(\sigma;t)=\frac{\sigma i^{\alpha}\sin\alpha\pi}{\pi}\int
_{0}^{\infty}\frac{e^{-xt}x^{\alpha-1}dx}{x^{2\alpha}-2\sigma\cos\alpha \pi%
\text{ }x^{\alpha}+\sigma^{2}},
\end{equation}
in closed form as 
\begin{equation}
F_{\alpha}(\sigma;t)=\frac{1}{\alpha\sigma t^{\alpha}}H_{2,3}^{2,1}\left(
\left. (-\sigma)^{1/\alpha}t\right\vert _{(\alpha,1),(0,1/\alpha
),(1-\alpha,-1)}^{(0,1/\alpha),(1-\alpha,-1)}\right) .
\end{equation}

The time dependent part of the wave function [Eq. (24)] can now be written
as 
\begin{equation}
T(t)=T(0)\left[ \frac{e^{i\lambda _{n}^{1/\alpha }t}}{\alpha }-\frac{1}{%
\alpha \sigma t^{\alpha }}H_{2,3}^{2,1}\left( \left. (-\sigma )^{1/\alpha
}t\right\vert _{(\alpha ,1),(0,1/\alpha ),(1-\alpha ,-1)}^{(0,1/\alpha
),(1-\alpha ,-1)}\right) \right] .
\end{equation}%
We now use Equation (C3) to write $F_{\alpha }(\sigma ;t)$ in computable
form as 
\begin{equation}
F_{\alpha }(\sigma ;t)=-\frac{1}{\alpha }\sum_{\nu =0}^{\infty }(-1)^{\nu (1+%
\frac{1}{\alpha })}\frac{\Gamma (-\frac{\nu }{\alpha })\Gamma (1+\frac{\nu }{%
\alpha })}{\Gamma (-\nu )\nu !}\frac{(\lambda _{n}i^{\alpha })^{\frac{\nu }{%
\alpha }}t^{\nu }}{\nu !}-\sum_{\nu =0}^{\infty }\frac{(\lambda
_{n}i^{\alpha })^{\nu }t^{\nu \alpha }}{\Gamma (1+\alpha \nu )}.
\end{equation}%
Since $\mu =1>0$, the above series converges for all $\left\vert
z\right\vert \neq 0.$ In the first series, we concentrate on the expression 
\begin{equation}
I=(-1)^{\nu (1+\frac{1}{\alpha })}\frac{\Gamma (-\frac{\nu }{\alpha })\Gamma
(1+\frac{\nu }{\alpha })}{\Gamma (-\nu )\nu !}.
\end{equation}%
We extend the formula 
\begin{equation}
\frac{\Gamma (-n)}{\Gamma (-N)}=(-1)^{N-n}\frac{N!}{n!},
\end{equation}%
where $N$ and $n$ are positive integers, to non integer arguments as 
\begin{equation}
\frac{\Gamma (-q)}{\Gamma (-Q)}=(-1)^{Q-q}\frac{\Gamma (Q+1)}{\Gamma (q+1)},
\end{equation}%
and write 
\begin{equation}
\frac{\Gamma (-\frac{\nu }{\alpha })}{\Gamma (-\nu )}=\mp (-1)^{\nu -\frac{%
\nu }{\alpha }}\frac{\Gamma (\nu +1)}{\Gamma (\frac{\nu }{\alpha }+1)}.
\end{equation}%
We have inserted the $\mp $ sign to display the ambiguity in the gamma
function for the negative integer values of its argument, where it diverges
as $\mp \infty $. Thus, $I$ is evaluated as 
\begin{align}
I& =\mp (-1)^{\nu (1+\frac{1}{\alpha })}(-1)^{\frac{\nu }{\alpha }-\nu }%
\frac{\Gamma (\nu +1)}{\Gamma (\frac{\nu }{\alpha }+1)}\frac{\Gamma (1+\frac{%
\nu }{\alpha })}{\nu !} \\
& =\mp 1.
\end{align}%
Substituting this into $F_{\alpha }(\sigma ;t)$ [Eq. (D28], we obtain 
\begin{align}
F_{\alpha }(\sigma ;t)& =\pm \frac{1}{\alpha }\sum_{\nu =0}^{\infty }\frac{%
(\lambda _{n}i^{\alpha })^{\frac{\nu }{\alpha }}t^{\nu }}{\nu !}-\sum_{\nu
=0}^{\infty }\frac{(\lambda _{n}i^{\alpha })^{\nu }t^{\nu \alpha }}{\Gamma
(1+\alpha \nu )} \\
& =\pm \frac{1}{\alpha }\sum_{\nu =0}^{\infty }\frac{(i\lambda
_{n}^{1/\alpha }t)^{\nu }}{\nu !}-\sum_{\nu =0}^{\infty }\frac{(\lambda
_{n}i^{\alpha }t^{\alpha })^{\nu }}{\Gamma (1+\alpha \nu )},
\end{align}%
which is nothing but 
\begin{equation}
F_{\alpha }(\sigma ;t)=\pm \frac{e^{i\lambda _{n}^{1/\alpha }t}}{\alpha }%
-E_{\alpha }(\lambda _{n}i^{\alpha }t^{\alpha }).
\end{equation}%
Substituting this into Equation (24), we write the time dependent part of
the wave function as 
\begin{equation}
T(t)=T(0)\left[ \frac{e^{i\lambda _{n}^{1/\alpha }t}}{\alpha }\mp \frac{%
e^{i\lambda _{n}^{1/\alpha }t}}{\alpha }+E_{\alpha }(\lambda _{n}i^{\alpha
}t^{\alpha })\right] .
\end{equation}%
To be consistent with the robust result in Equation (21), we pick the minus
sign. Thus the time dependent part of the wave function is again obtained as 
\begin{equation}
T(t)=E_{\alpha }(\lambda _{n}i^{\alpha }t^{\alpha }),
\end{equation}%
where without loss of any generality we have set $T(0)=1.$

\end{document}